# Phase-controlled Fourier-transform spectroscopy

Kazuki Hashimoto[1,2] and Takuro Ideguchi[1,3]*

[1]Department of Physics, The University of Tokyo, Tokyo 113-0033, Japan
[2]Aeronautical Technology Directorate, Japan Aerospace Exploration Agency, Tokyo 181-0015, Japan
[3]PRESTO, Japan Science and Technology Agency, Tokyo 113-0033, Japan
*ideguchi@phys.s.u-tokyo.ac.jp

**Fourier-transform spectroscopy (FTS) has been widely used as a standard analytical technique over the past half-century. FTS is a simple and robust autocorrelation-based technique that is compatible with both temporally coherent and incoherent light sources, which functions as an active or passive spectrometer. However, this technique has been mostly used for static measurements due to the low scan rate imposed by technological restrictions. This has impeded its application to continuous rapid measurements, which would be of significant interest for a variety of fields, especially when monitoring of non-repeating/transient complex dynamics is desirable. Here, we demonstrate highly efficient FTS operating at a high spectral acquisition rate with a simple delay line based on a dynamic phase-control technique. The independent adjustability of phase and group delays allows us to achieve the Nyquist-limited spectral acquisition rate over 10,000 spectra per second, while maintaining a large spectral bandwidth and high resolution. In addition, we demonstrate the ability of this passive spectrometer working with an incoherent light source.**

Fourier-transform spectroscopy (FTS) has provided solutions to countless chemical analysis problems in the fields of chemistry, pharmacy, medicine, biology and physics including material science[1-3]. It enables the simultaneous identification of multiple molecular species via high-resolution broadband molecular spectra measured by a simple optical system consisting of a scanning Michelson interferometer and a broadband light source. The fact that this analytical method has long been widely used owes to the simple and robust autocorrelation-based working principle, enabling a variety of uses. For example, since it allows us to measure temporally incoherent light, FTS can be used as a passive spectroscopy technique with an external light source such as sunlight[4,5]. However, FTS has been mostly used for measuring static samples because of the low temporal resolution due to the low scan rate of mechanical delay lines. This is incompatible for investigating fast phenomena at a higher temporal resolution, such as combustion processes which often require a measurement rate over 10 kHz[6,7]. Rapid-scan FTS instruments typically operate at an improved temporal resolution only at a low spectral resolution, for example a scan rate of 667 Hz with a resolution of 120 GHz (4 cm$^{-1}$)[8] or 77 kHz with a resolution of 360 GHz (12 cm$^{-1}$)[9]. For a higher resolution applicable to gas-phase analysis such as 15 GHz (0.5 cm$^{-1}$), the available scan rate significantly drops down to less than 10 Hz[10,11] due to the lack of a rapid-scan long-range delay line. Step-scan FTS is also known to improve the temporal resolution but it works only for measuring repeatable phenomena, which significantly limits applications.

Recent development in dual-comb spectroscopy (DCS)[12,13], a FTS technique based on mutually coherent laser frequency combs with slightly detuned pulse repetition rates, has elegantly solved the problem and shown a tremendous improvement on the measurement scan rate[14-31]. The superiority of this technique is its highest scan rate, i.e., Nyquist-limited scan rate, enabled by fully utilizing the Nyquist range, determined by half a repetition rate of the laser. In other words, DCS has been a unique FTS that enables the highest measurement rate, spectral bandwidth and resolution to be achieved under the Nyquist-limited trade-off constraint. This capability owes to the adjustable difference in both the repetition rate and carrier-envelope offset (CEO) frequency between the combs. The former determines the down conversion factor from the optical to radio frequency (RF) to achieve the highest possible scan rate, while the latter adjusts the position of the RF spectrum within the Nyquist range for avoiding aliasing. The latter capability has not often been clearly appreciated in the previous reports on dual-comb spectroscopy, but we emphasize that it assures DCS to be an efficient FTS technique. In return for this superior capability, DCS requires a more elaborate and sophisticated system than the conventional Michelson-type FTS because DCS is based on the cross-correlation measurement between the two lasers, for which mutual coherence is usually fragile. Furthermore, it is an active-type spectrometer and not suitable to be used in a passive manner[17,19].

In this work, we revisit conventional Michelson-type FTS and develop a Nyquist-limited highly efficient FTS, called phase-controlled FTS (PC-FTS), that allows us to arbitrarily adjust the scan rate, spectral bandwidth, and spectral resolution under a tradeoff relation among them. This approach uses a rapid-scan phase-control mechanism inspired by pulse shaping techniques[32], that enables us to acquire interferograms continuously at a rate beyond 10 kHz. This is the highest scan rate achievable as a Michelson-type FTS, while avoiding spurious spectral distortions due to the aliasing effect. This is made possible by the arbitrary and independent adjustment of the group and phase delay added by the delay line, which can be understood as an analogy of the independent adjustment of the difference in repetition rate and CEO frequency in a dual-comb spectrometer. As a proof of concept demonstration, we measure broadband gas absorption spectra of hydrogen cyanide ($H^{12}C^{14}N$) and acetylene ($^{12}C_2H_2$) in the near-infrared region spanning over 1.8 THz with a resolution of 11.5 GHz at an acquisition rate of over 12 kHz both with a temporally coherent and incoherent light sources. Our system holds great promise for applications aimed at measuring non-repeating phenomena at high-temporal-resolution such as a transient combustion process, or a large number of events at high-throughput such as large-area environmental monitoring.

## Results
**Principle of phase-controlled Fourier-transform spectroscopy.**

A schematic of PC-FTS is shown in Fig. 1**a**. It consists of a broadband light source, a scanning Michelson interferometer, a single photodetector and a digitizer. Since Michelson-type FTS is an autocorrelation-based measurement, there is no requirement on the temporal coherence for the measured light source. In our experiment, a coherent mode-locked laser or an incoherent amplified spontaneous emission from a super-luminescent diode is used as a broadband light source. The scanning Michelson interferometer varies a time delay between the beams traveling along the reference and scan arms, so that an autocorrelation trace, called the interferogram, is obtained as a function of time by the photodetector. The digitized interferogram is Fourier-transformed to obtain a spectrum. By placing a sample in the optical path between the interferometer and the detector, its absorption feature can be encoded onto the spectrum. Our phase-controlled delay line consists of a dispersive element such as a diffraction grating, a focusing element and a scanning mirror, which are aligned in a 4f geometry such that each spectral component of the beam is focused at a different spot on the Fourier plane, where the scanning mirror is placed[33,34]. The reflected beam from the scanning mirror then passes through the optical elements in the 4f system and is retro-reflected by an end mirror. The 4f geometry ensures the beam travels back along the incident path for any facet-angles of the scanning mirror. Using a rapidly scanning mirror enables the system to function as a rapid-scan FTS. A detailed schematic of the PC-FTS is shown in Fig. 2 and the components used in the system are described in Methods section.

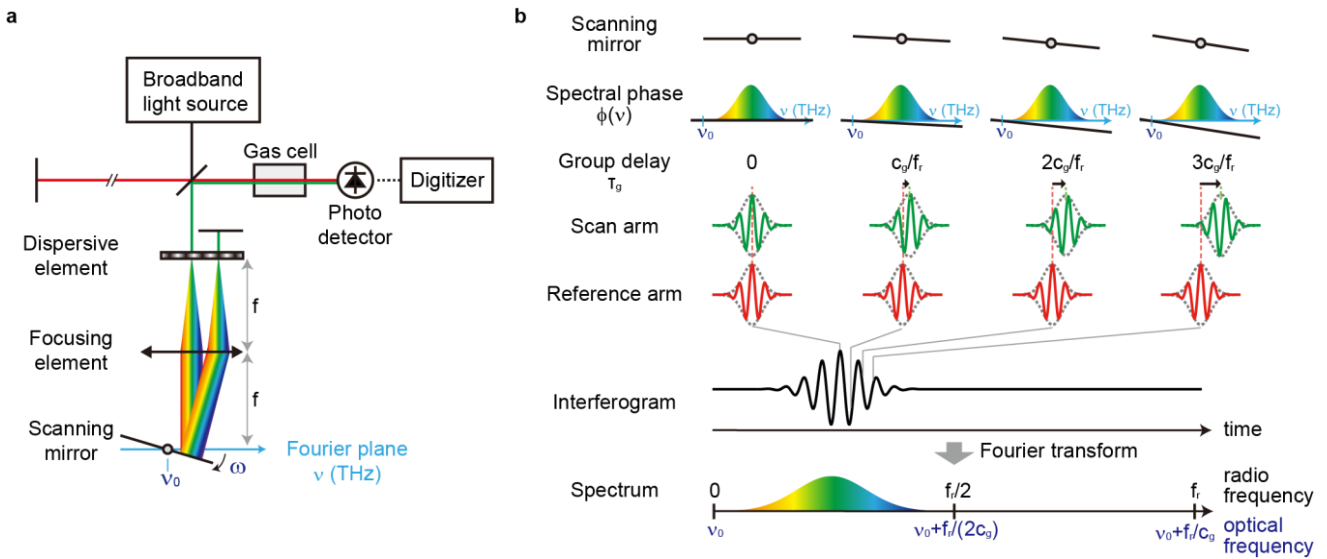

**Fig. 1| Schematic and conceptual illustration of phase-controlled Fourier-transform spectrometer. a**, Schematic of the system. The system consists of a broadband light source, a scanning Michelson interferometer with the rapid-scan phase-controlled delay line, a sample gas cell, a photodetector and a digitizer. The delay line consists of a dispersive element, a focusing element, and a scanning mirror aligned in a reflective 4f-configuration. The broadband light is focused at the Fourier plane in the 4f system after spectral separation by the dispersive element such that each spectral component is mapped on a different position at the Fourier plane. The scanning mirror changes its angle at an angular frequency of $\omega$ and reflects the light with angled directions. The beam traveling through the 4f system is retro-reflected with an end mirror and goes back along the same path. The corresponding optical frequency of the pivot position of the scanning mirror in the Fourier plane is indicated as $\nu_0$. **b**, Concept of PC-FTS. The upper part shows how the interferogram is obtained in the time domain. For simplicity, the light fields are depicted as a pulse train with zero carrier-envelope-phase as shown in the pulses in the reference arm. At each time frame, the light field in the scan arm acquires a linear spectral phase that is proportional to the angle of the scanning mirror in the delay line. The linear spectral phase ramp adds a group delay and phase delay to the light field. The pivot frequency $\nu_0$ is converted to the zero frequency in the down-converted RF spectrum after Fourier-transformation. Therefore, the down-converted spectrum can be positioned inside of the Nyquist range by adjusting $\nu_0$.

The working principle of PC-FTS is depicted in Fig. 1**b**. A slight inclination of the scanning mirror placed at the Fourier plane adds a linear spectral phase ramp to the broadband light that travels through the delay line. Here, we describe the system with a galvanometric scanner. Assuming the scanning mirror rotates clockwise at a constant angular frequency $\omega$, the added spectral phase $\phi(\nu)$ and group delay $\tau_g$ may be described as:

$$\phi(\nu) = -2\pi c_g (\nu - \nu_0) t \tag{1}$$

$$\tau_g = -\frac{1}{2\pi}\frac{\partial \phi(\nu)}{\partial \nu} = c_g t \tag{2}$$

where $t$ denotes time, $\nu$ optical frequency, $\nu_0$ the optical frequency corresponding to the pivot position of the scanning mirror in the Fourier plane, and $c_g \propto \omega/\nu_0$ the down conversion factor, which is approximately a constant value in time. A full description of $c_g$ is described in Supplementary Information. The group delay is linearly proportional to time by a factor of $c_g$, while the group velocity can be described as $c_g c$, where $c$ denotes the speed of light. This indicates that a high angular frequency of the scanning mirror makes possible a large bandwidth of the down-converted RF spectrum, leading to an efficient use of the Nyquist range. Adjusting the angular frequency of the scanning mirror, which determines the down conversion factor of FTS, is analogous to adjusting the difference in repetition rate between the combs in DCS. Equation (1) tells us that a down-converted RF spectrum has a frequency of $c_g(\nu - \nu_0)$, that can be shifted by changing the pivot position of the scanning mirror $\nu_0$. Adjusting $\nu_0$ allows us to fully utilize the Nyquist range while avoiding the aliasing, which is analogous function as that of the difference in CEO frequency between the combs in DCS.

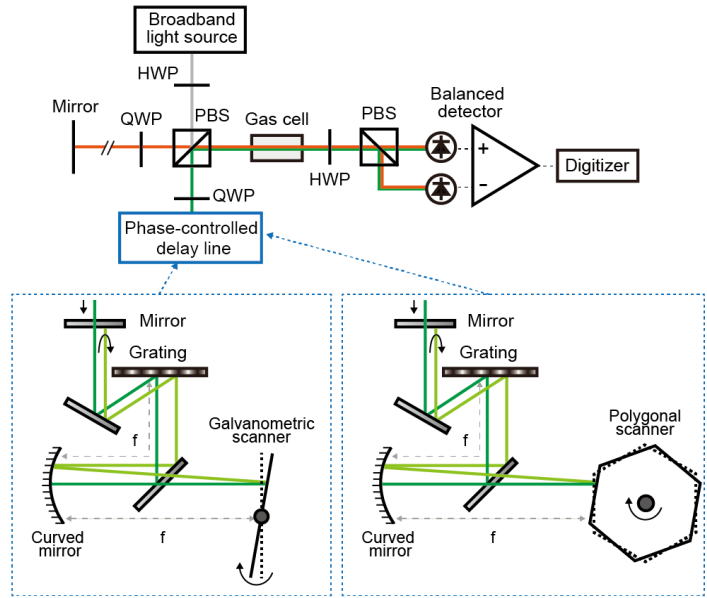

**Fig. 2| Detailed schematic of the phase-controlled FTS.** HWP: Half-wave plate, QWP: Quarter-wave plate, PBS: Polarizing beamsplitter.

**Phase and group delay correction.**
Since the down-conversion factor $c_g$ is not exactly a constant value in time, the phase-controlled delay lines generate temporally nonlinear phase and group delays, which must be corrected for FTS. The detailed description of the nonlinearity is provided in Supplementary Information. To linearize both the nonlinear phase and group delays, we use continuous-wave (CW) interferograms at two different optical frequencies and follow the correction procedure that is in principle the same as that of DCS[16,35]. Note that the fixed instrumental nonlinearities of the phase and group delays are corrected in the PC-FTS, while the random temporal fluctuations of the difference in repetition rate and CEO frequency are corrected in the DCS. Since the PC-FTS is an autocorrelation-based technique where random temporal fluctuation is inherently small, the correction interferograms are not necessarily measured for every measurement.

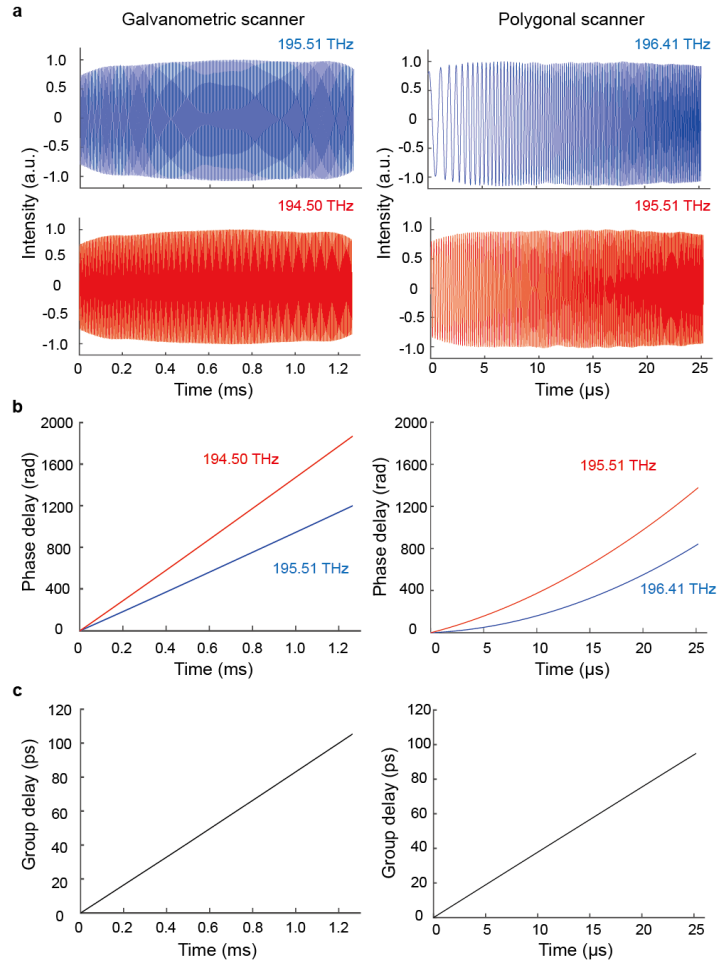

**Fig. 3| Characterization of the phase and group delays added by the delay lines. a,** Measured continuous-wave interferograms at the two different frequencies. **b,** Phase delays of the CW interferograms. **c,** Group delays evaluated from the phase delays at the two different frequencies.

We measure the CW interferograms at two different frequencies with external-cavity laser diodes with a linewidth of less than 500 kHz lasing at 195.51 and 194.50 THz for the system with a galvanometric scanner, and 196.41 and 195.51 THz for the system with a polygonal scanner. The lasers' frequencies are measured by an optical spectrum analyzer (AQ6317B, Yokogawa) with the accuracy of ±2.6 GHz. The measured CW interferograms are shown in Fig. 3**a**. The phase nonlinearity is seen especially in the data measured by the system with the polygonal scanner. The phase delays are extracted from the interferograms and plotted in Fig. 3**b**. The group delays are calculated from the phase delays at the two different frequencies as shown in Fig. 3**c**. The maximum group delay is about 100 ps, which corresponds to the spectral resolution of about 10 GHz.

**Broadband absorption spectroscopy with a galvanometric scanner.**

As a proof-of-concept demonstration of the PC-FTS, we measure broadband absorption spectra of gaseous HCN by the system with a galvanometric scanner. An Er-fibre mode-locked laser that generates femtosecond pulses at a repetition rate of 50 MHz is used as a light source. A full description of the system is provided in Methods section. The interferograms shown in Fig. 4**a** display the clear modulations of the molecular free-induction decay. By Fourier-transforming a single-sided interferogram with the aforementioned phase and group delay correction[16,35] and the conventional phase correction methods[36-40], a broadband spectrum spanning over 7 THz is obtained at a resolution of 10.1 GHz with a high signal-to-noise ratio (SNR) (Fig. 4**b**). Sharp absorption lines of the overtone vibrational bands of HCN are clearly observed. Note that the resolution of 10 GHz corresponds to the maximum optical path length difference of 30 mm, which is achieved only by tilting the scanning mirror. The scan rate is 300 Hz, which is not limited by the Nyquist range (0-25 MHz) determined by the pulse repetition rate of the mode-locked laser but the scan rate of the galvanometric scanner.

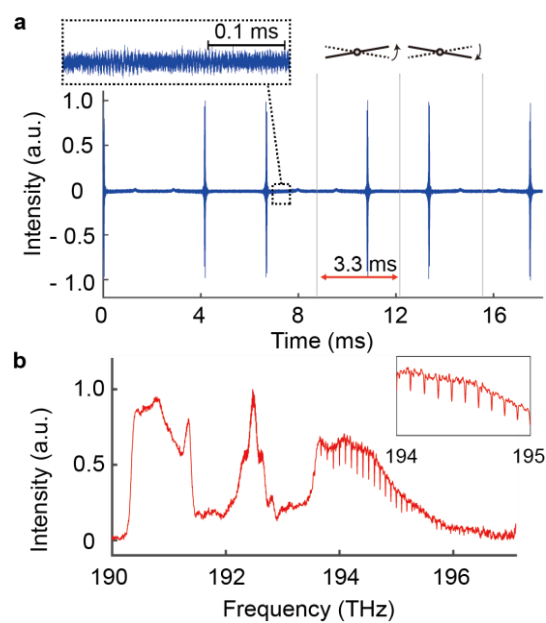

**Fig. 4| Broadband absorption spectroscopic measurement of $H^{12}C^{14}N$ molecules by PC-FTS based on a galvanometric scanner.**
**a**, Continuous interferograms of $H^{12}C^{14}N$ molecules measured with the mode-locked laser. The single-sided interferograms are continuously obtained at a scan rate of 300 Hz (the corresponding temporal interval is 3.3 ms). The inset shows modulations due to the free-induction decay. **b**, A broadband spectrum obtained by Fourier-transforming a single interferogram. The non-averaged broadband spectrum covering 7 THz shows absorption lines of HCN molecules at a spectral resolution of 10.1 GHz with a high signal-to-noise ratio. The structured spectral profile comes from that of the mode-locked fibre laser.

**Nyquist-limited rapid-scan spectroscopy with a polygonal scanner.**

To fully utilize the Nyquist range and achieve the Nyquist-limited highest scan rate, a polygonal scanner can be used instead of the galvanometric scanner. With the faster rotation speed of the polygonal scanner, a continuous measurement of interferograms at a rate of 12 kHz is demonstrated in Fig. 5a. Figure 5b shows the corresponding consecutive broadband $C_2H_2$ absorption spectra with a bandwidth of 1.5 THz at a spectral resolution of 11.5 GHz. The spectra are normalized and converted to transmittance by a baseline fitting with Savitzky–Golay filtering[41]. Note that the effective scan velocity is 312 m/s, which is orders of magnitude higher than those of conventional delay lines. Figure 5c shows a 20-averaged spectrum compared to a computationally retrieved spectrum based on HITRAN database[42]. The SNR of a single spectrum is 54, which is limited by detector noise in this proof-of-concept demonstration. It could be improved by using a detector with better noise performance.

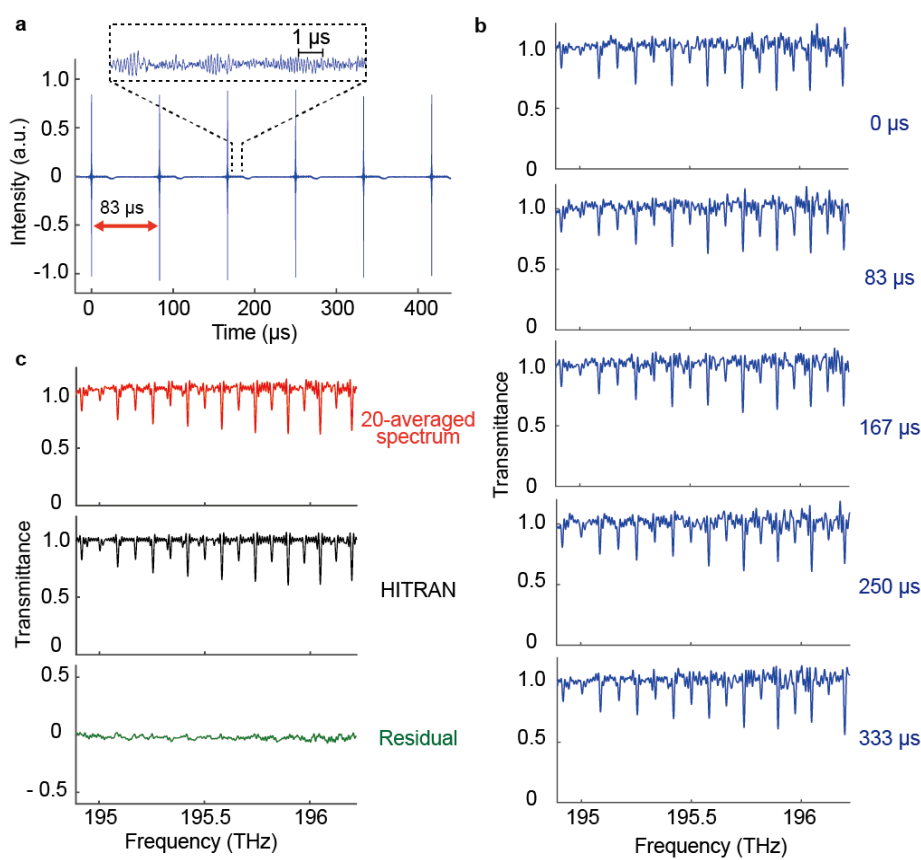

**Fig. 5| High-scan-rate broadband absorption spectroscopic measurement of $^{12}C_2H_2$ molecules by PC-FTS based on a polygonal scanner. a**, Continuously measured interferograms with the mode-locked laser. The burst of the interferogram appears every 83 μs, corresponding to a scan rate of 12 kHz. Clear modulations of the molecular free-induction decay are observed. **b**, Non-averaged transmission spectra corresponding to the interferograms shown in Fig. 5a. Each spectrum covers over 1.5 THz with a resolution of 11.5 GHz. The absorption lines of $C_2H_2$ are clearly observed. The noise in the spectra can mainly be attributed to the side-modulations of the sinc function caused by the rectangular Fourier-transform windowing. **c**, Comparison between a 20-averaged transmission spectrum and a computationally calculated spectrum based on HITRAN database. The standard deviation of the residual is 1.7%.

**PC-FTS with an incoherent light source.**

Since the PC-FTS is an autocorrelation-based technique, it functions as a passive spectrometer with a temporally incoherent light source. Furthermore, since a repetition rate is not usually defined for an incoherent light, the Nyquist frequency of the measurement is determined by an electronic sampling rate of a digitizer. This enables a higher scan rate with a larger Nyquist range.

To show the applicability of an incoherent light to the PC-FTS, we replace the mode-locked laser to a commercially available super-luminescent diode (SLD) and perform a proof-of-concept measurement. We acquire HCN absorption spectra over 1.8 THz with a spectral resolution of 11.5 GHz at a scan rate of 24 kHz, which is higher than that of the mode-locked laser by a factor of two. Note that the scan rate of 24 kHz is not achievable with the mode-locked laser at the repetition rate of 50 MHz because of the aliasing effect caused by the Nyquist constraint due to the pulse repetition rate. Measured continuous interferograms are shown in Fig. 6**a**, where free-induction decay of the molecules is clearly seen every 42 μs. A spectrum Fourier-transformed from a single interferogram and a 30-averaged spectrum are shown in Fig. 6**b**. The fringe observed in each spectrum is the ripple-noise due to the multiple-reflections between the facets of the SLD chip.

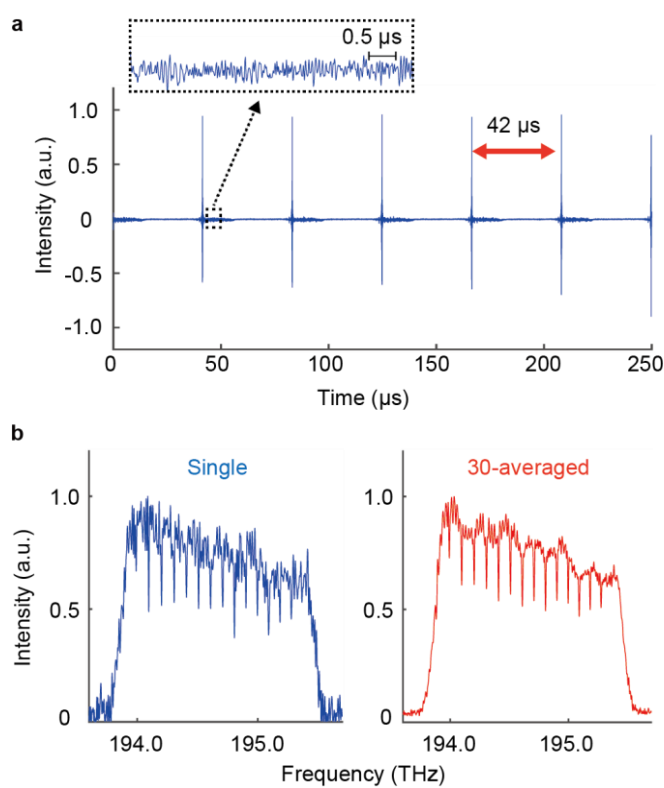

**Fig. 6| Interferograms and spectra measured by the PC-FTS with the SLD. a,** Continuous interferograms of $H^{12}C^{14}N$ molecules measured with the SLD at a scan rate of 24 kHz. **b**, Spectra obtained by Fourier-transforming a single interferogram (left), and a 30-averaged spectrum.

**Signal-to-noise ratio.**

We evaluate SNR of the PC-FTS. Dominant noises are categorized as the detector noise, shot noise and relative intensity noise (RIN) of the light source[43]. In our experiment, an average power irradiated onto the balanced detector is 15 μW for each photodiode, which is limited by the detector nonlinearity. From the noise equivalent power (NEP) of our balanced detector (ca. 4.7 pW/$\sqrt{Hz}$), the SNR dominated by the detector noise is estimated to be 82, while that of the shot noise to be 130. From the measured RIN of our mode-locked laser (-143 dB/Hz), the SNR dominated by the RIN is estimated to be 255. Therefore, the SNR of our measurement is dominated by the detector noise, and using a detector with a lower NEP could increase the SNR. The overall SNR including all the noises is evaluated to be 66, which is in good agreement with the experimentally measured value of 54. In the case of the measurement with the SLD, relatively larger RIN dominates the SNR. The measured RIN of our SLD is -124 dB/Hz, that leads to the SNR of 29. In addition, the SLD used in our experiment generates a strong ripple noise due to the multiple reflections on the chip surfaces. However, the ripple noise is not a fundamental noise associated to the PC-FTS, which can be eliminated with a proper surface treatment of the chip.

**Discussion**

The concept of this technique can be applied for other wavelengths including the mid-infrared region, where the normal vibrational modes of molecules exist. The single-photodetector-operation of the FTS is especially advantageous in the mid-infrared region because of the lack of a high-quality mid-infrared detector-array required for a high-speed dispersive spectrometer. Furthermore, PC-FTS is useful not only for gas-phase molecules but also for liquid- or solid-phase materials. As discussed above, since we can arbitrarily choose a combination of the scan rate, spectral bandwidth and resolution, a broader spectral bandwidth, for example, can be measured by reducing the spectral resolution while keeping the scan rate. Finally, this highly efficient FTS technique can also be modified and applied to multi-dimensional Fourier-transform spectroscopy, which requires time-consuming scans of multiple delay lines[29].

**Methods**

**Phase-controlled delay line.**

Our phase-controlled delay line consists of a dispersive element, a focusing element, a scanning mirror, and a flat mirror. These components are aligned to form a reflective 4f configuration such that a spectrum of the incident light is focused onto the Fourier plane where the scanning mirror is placed. In our demonstration, a reflective ruled grating with 600 lines/mm and a curved mirror with a focal length of 150 mm are used as a dispersive element and a focusing element, respectively. The number of grooves of the grating and the focal length of the curved mirror are to be set for optimizing the system (See Supplementary Information for details). A galvanometric scanner or a polygonal scanner with 36 facets are used as the scanning mirror. The mirror facets of both scanners are coated with gold. The facet size of the galvanometric scanner is 10 mm, and that of the polygonal scanner is 5.6 mm per each facet. The inner diameter of the polygonal scanner is 63.6 mm and its rotation speed is set to be 20,000 or 40,000 rpm (rotations per minute) in our experiments.

**Phase-controlled Fourier-transform spectroscopy.**

A full description of our PC-FTS is shown in Fig. 2. A temporally coherent mode-locked Er-doped fibre laser (Femtolite CS-20-GS, IMRA) or an incoherent super-luminescent diode (S5FC1005S, Thorlabs) is used as a broadband light source. The broadband light is delivered with a single-mode fibre to a fibre-collimator placed at the input port of the spectrometer. For a rapid measurement with the polygonal scanner, the light is optically band-pass-filtered with a span of 14 nm (corresponding to 1.8 THz) centered at 1533.5 nm (FBH1550-12, Thorlabs). The collimated light is split into two beams with a half-wave plate and a polarizing beamsplitter. The rotation angle of the half-wave plate is set to have a power ratio of 50:50 between the two output beams of the interferometer. While the beam in the reference arm is retro-reflected by a flat mirror, the other beam in the scan arm travels through the phase-controlled delay line and retro-reflected with a time delay. A quarter-wave plate inserted in each arm rotates the polarization of the reflected beam by 90 degrees, so that the recombined beams exit from the output port of the interferometer with orthogonal polarizations. The beams go through a gas cell containing either $H^{12}C^{14}N$ (TT-HCN-10OT-G-Q, triad technology) or $^{12}C_2H_2$ (TT-CH12-50T-G-Q, triad technology) in a two-way (round trip) geometry and detected by an InGaAs balanced photodetector (PDB415C-AC, Thorlabs). For balancing the photodiodes' signals, a half-wave plate and a polarizing beamsplitter are used in front of the detector. The average power of the detected light is set to be 15 µW for each photodiode to avoid effects of detector nonlinearity. The detector signal is lowpass-filtered at 21 MHz when using the mode-locked laser. The signal is digitized by a data acquisition board (ATS9440, AlazarTech) at a sampling rate of 125 MS/s. The digitized time domain data is segmented into independent interferograms and each of them is Fourier-transformed after the phase correction.

**Acknowledgement**

We thank Makoto Kuwata-Gonokami and Junji Yumoto for letting us use their equipment. We are grateful to Faris Sinjab, Venkata Ramaiah Badarla, and Miu Tamamitsu for their critical reading of the manuscript. This work was financially supported by JSPS KAKENHI (17H04852, 17K19071), JST PRESTO (JPMJPR17G2), Advanced Photon Science Alliance (APSA), Research Foundation for Opto-Science and Technology, and The Murata Science Foundation.

**Author contributions**

TI conceived the work. KH and TI designed the system. KH conducted the experiment and analyzed the data. TI supervised the work. KH and TI wrote the manuscript.

# Supplementary Information: Phase-controlled Fourier-transform spectroscopy

## A. Theoretical description of the phase-controlled FTS

### A-1. Phase-controlled delay line with a galvanometric scanner

Theoretical description of the phase-controlled delay line is provided as follows. The detailed schematic of the phase-controlled delay line with a galvanometric scanner is depicted in Fig. S1**a**. In this configuration, the grating equation may be described as:

$$\sin\beta_\lambda = N\lambda - \sin\alpha, \quad (S1)$$

$$\sin\beta_{\lambda_0} = N\lambda_0 - \sin\alpha, \quad (S2)$$

where $\beta_\lambda$ and $\beta_{\lambda_0}$ denote the first-order diffraction angles of the light at wavelength of $\lambda$ and $\lambda_0$, $\alpha$ incident angle, $N$ groove density of the grating. Here, $\lambda_0$ denotes the wavelength corresponding to the pivot position of the scanning mirror in the Fourier plane. Assuming the facet of the scanning mirror is on the Fourier plane at $t = 0$, an optical path length difference $L_g(t)$ between the beams at the wavelength of $\lambda$ and $\lambda_0$ is described as:

$$L_g(t) = l_f(\tan\beta_{\lambda_0} - \tan\beta_\lambda)\tan\omega t, \quad (S3)$$

where $l_f$ denotes the focal length of the curved mirror and $\omega$ the angular frequency of the scanning mirror. When $\beta_{\lambda_0}, \beta_\lambda \ll 1$ and $\omega t \ll 1$, $L_g(t)$ is described as:

$$L_g(t) \approx l_f N(\lambda_0 - \lambda)\omega t$$
$$\approx l_f N c \left(\frac{\nu - \nu_0}{\nu_0 \nu}\right)\omega t, \quad (S4)$$

where $c$ is the speed of light, $\nu$ and $\nu_0$ the optical frequency corresponding to $\lambda$ and $\lambda_0$. Then, the spectral phase $\phi_g(\nu)$ may be described as:

$$\phi_g(\nu) = -\frac{2\pi\nu}{c} \times 4 \times L_g(t)$$
$$\approx -2\pi \frac{4l_f N\omega}{\nu_p}(\nu - \nu_0)t \quad (S5)$$
$$= -2\pi c_g(\nu - \nu_0)t,$$

where $c_g = \frac{4l_f N\omega}{\nu_0}$ is the down conversion factor. Since the light is reflected off the scanning mirror twice in the delay line, the total optical path length difference is $L_g(t)$ multiplied by 4. The group delay $\tau_g$ may be described as:

$$\tau_g = -\frac{1}{2\pi}\frac{\partial \phi_g(\nu)}{\partial \nu} = c_g t. \quad (S6)$$

### A-2. Phase-controlled FTS with a galvanometric scanner

We denote complex electric fields of an input and two outputs of the Michelson interferometer as $E(t)$, $E_{scan}(t)$ and $E_{ref}(t)$, and their Fourier-transformed fields as $E(\nu)$, $E_{scan}(\nu)$ and $E_{ref}(\nu)$, respectively. In the frequency domain, $E_{ref}(\nu)$ and $E_{scan}(\nu)$ may be described by adding the spectral phase to $E(\nu)$ as:

$$E_{ref}(\nu) = E(\nu)\exp(-i2\pi\nu\tau), \tag{S7}$$

$$\begin{aligned}E_{scan}(\nu) &= E(\nu)\exp\left[i\left\{-2\pi\nu\tau + \phi_g(\nu)\right\}\right] \\ &= E(\nu)\exp[-i2\pi\{\nu\tau + (\nu - \nu_0)\tau_g\}],\end{aligned} \tag{S8}$$

where $\tau$ denotes a delay added by the reference and scan arm of the interferometer, and $\tau_g$ the group delay added by the delay line in the scan arm. By inverse-Fourier-transforming $E_{ref}(\nu)$ and $E_{scan}(\nu)$, the complex electric fields $E_{ref}(t)$ and $E_{scan}(t)$ are described as:

$$\begin{aligned}E_{ref}(t) &= \mathcal{F}^{-1}\{E_{ref}(\nu)\} \\ &= \int_{-\infty}^{\infty} E(\nu)\exp\{i2\pi\nu(t-\tau)\}d\nu \\ &= E(t-\tau),\end{aligned} \tag{S9}$$

$$\begin{aligned}E_{scan}(t) &= \mathcal{F}^{-1}\{E_{scan}(\nu)\} \\ &= \exp(i2\pi\nu_0\tau_g)\int_{-\infty}^{\infty} E(\nu)\exp\{i2\pi\nu(t-\tau-\tau_g)\}d\nu \\ &= \exp(i2\pi\nu_0\tau_g)E(t-\tau-\tau_g).\end{aligned} \tag{S10}$$

Integrated intensity of the combined electric fields of $E_{ref}(t)$ and $E_{scan}(t)$ detected by a photodetector can be written as:

$$\begin{aligned}I(\tau_g) &= \int_{-\infty}^{\infty} |E_{ref}(t) + E_{scan}(t)|^2 dt \\ &= \int_{-\infty}^{\infty} |E(t) + \exp(i2\pi\nu_0\tau_g)E(t-\tau_g)|^2 dt.\end{aligned} \tag{S11}$$

Here, $t - \tau$ is replaced by $t$ for simplicity of the equation. The AC component of $I(\tau_g)$, namely the interferogram $S(\tau_g)$, is described as:

$$S(\tau_g) = \exp(i2\pi\nu_0\tau_g) \int_{-\infty}^{\infty} E^*(t)E(t-\tau_g)dt + c.c.. \tag{S12}$$

Finally, Fourier-transforming the interferogram gives the spectrum:

$$\begin{aligned}\mathcal{F}\{S(\tau_g)\} &= \int_{-\infty}^{\infty} S(\tau_g)\exp(-i2\pi\nu\tau_g)d\tau_g \\ &= \int_{-\infty}^{\infty} \exp\{-i2\pi(\nu-\nu_0)\tau_g\}s(\tau_g)d\tau_g + c.c. \\ &= B(\nu-\nu_0) + c.c..\end{aligned} \tag{S13}$$

Here, $s(\tau_g) = \int_{-\infty}^{\infty} E^*(t)E(t-\tau_g)dt$ and $B(\nu) = \mathcal{F}\{s(\tau_g)\}$. As shown above, the spectrum is shifted by $\nu_0$, which is experimentally demonstrated by changing the pivot position of the scanning mirror in the Fourier plane.

## A-3. Phase-controlled FTS with a polygonal scanner

Unlike a galvanometric scanner, a polygonal scanner does not have the pivot in the Fourier plane (Fig. S1**b**). Therefore, the delay line based on the polygonal scanner is described in a different manner. In this configuration, the optical path length difference $L_p(t)$ added by the delay line may be described as:

$$L_p(t) \approx \left\{ l_f N (\lambda_0 - \lambda) - R \tan\frac{\omega t}{2} \right\} \tan \omega t$$

$$\approx l_f N c \left( \frac{\nu - \nu_0}{\nu_0 \nu} \right) \omega t - R \frac{\omega^2 t^2}{2}, \tag{S14}$$

where $R$ denotes the inner radius of the polygonal scanner that equals to the distance between the pivot and the mirror facets. The added spectral phase $\phi_p(\nu)$ may be described as:

$$\phi_p(\nu) = -\frac{2\pi\nu}{c} \times 4 \times L_p(t)$$

$$\approx -2\pi \frac{4 l_f N \omega}{\nu_p}(\nu - \nu_0) t + 2\pi \frac{2 R \omega^2 \nu}{c} t^2$$

$$= -2\pi \left\{ \left( \frac{4 l_f N \omega}{\nu_0} - \frac{2 R \omega^2}{c} t \right) \nu - 4 l_f N \omega \right\} t \tag{S15}$$

$$= -2\pi \left( \frac{4 l_f N \omega}{\nu_0} - \frac{2 R \omega^2}{c} t \right) \left( \nu - \frac{\nu_0}{1 - \frac{R \omega \nu_0}{2 c l_f N} t} \right) t$$

$$= -2\pi c_p (\nu - \nu'_0) t,$$

where $c_p = \left( \frac{4 l_f N \omega}{\nu_0} - \frac{2 R \omega^2}{c} t \right)$ denotes the down conversion factor and $\nu'_0 = \frac{\nu_0}{1 - \frac{R \omega \nu_0}{2 c l_f N} t}$ the optical frequency corresponding to the intersection point of the scanner's facet on the Fourier plane. Note that the above expressions become identical to those with the galvanometric scanner shown in the previous section with $R = 0$. The group delay $\tau_p$ may be described as:

$$\tau_p = -\frac{1}{2\pi} \frac{\partial \phi_p(\nu)}{\partial \nu} = c_p t. \tag{S16}$$

In the frequency domain, $E_{scan}(\nu)$ may be described as:

$$E_{scan}(\nu) = E(\nu) \exp\{-i2\pi(\nu - \nu'_0)\tau_p\}, \tag{S17}$$

Here, we omit $\tau$ for simplicity of the equation. Then, $E_{scan}(t)$ is written as:

$$E_{scan}(t) = \exp(i2\pi\nu'_0 \tau_p) \int_{-\infty}^{\infty} E(\nu) \exp\{i2\pi\nu(t - \tau_p)\} d\nu$$

$$= \exp(i2\pi\nu'_0 \tau_p) E(t - \tau_p). \tag{S18}$$

Therefore, the interferogram $S(\tau_p)$ is written as:

$$S(\tau_p) = \int_{-\infty}^{\infty} E^*(t) E(t - \tau_p) \exp(i2\pi\nu'_0 \tau_p) dt + c.c.. \tag{S19}$$

Since $\tau_p$ and $\nu'_0$ are temporally nonlinear variables, phase correction is necessary to retrieve the spectrum.

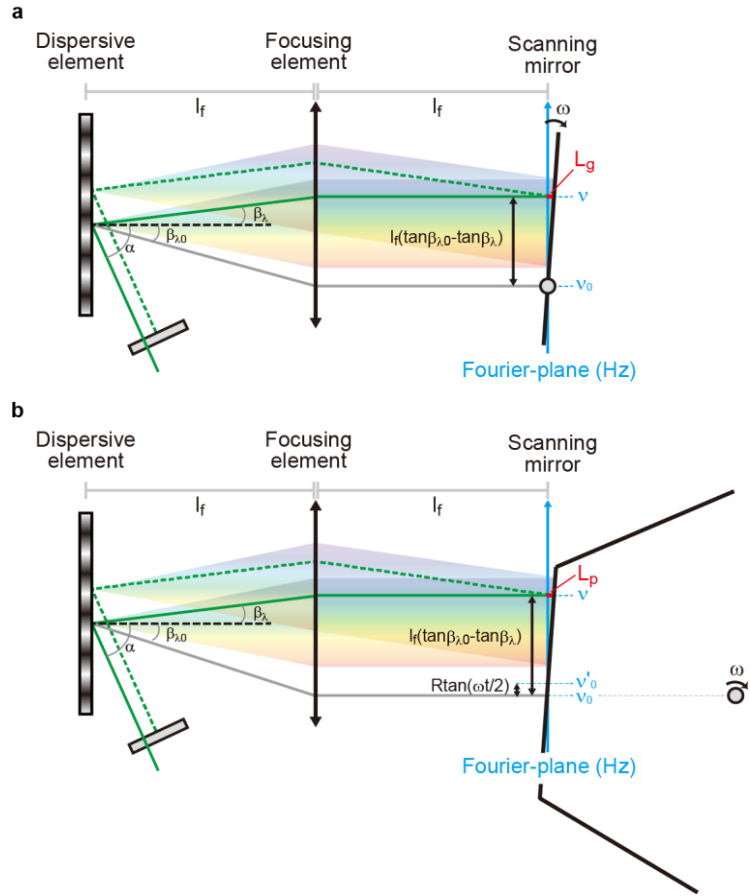

**Fig. S1| Detailed schematic of the phase-controlled delay line. a,** Schematic of the phase-controlled delay line with a galvanometric scanner. **b,** Schematic of the phase-controlled delay line with a polygonal scanner.

## B. Trade-off relation among the scan rate, spectral bandwidth and resolution
### B-1. Phase-controlled FTS with a galvanometric scanner

A trade-off relation among the scan rate, spectral bandwidth and resolution of the phase-controlled FTS with a galvanometric scanner can be derived as follows. A down-converted radio frequency $f(\nu)$ may be described as:

$$f(\nu) = \frac{4l_f N\omega}{\nu_0}(\nu - \nu_0) = c_g(\nu - \nu_0), \tag{S20}$$

where $c_g$ denotes the down conversion factor, $\nu_0$ the optical frequency corresponding to the pivot position of the scanning mirror in the Fourier plane. All the spectral elements $f(\nu)$ of the measured light must be within the Nyquist range of the system to avoid the aliasing effect:

$$0 \leq f(\nu) < \frac{f_s}{2}, \tag{S21}$$

where $f_s$ is a sampling rate of the system, which can be determined by either an optical or electrical sampling rate, namely the repetition rate of the pulsed laser or the clock rate of the digitizer, respectively. For fully utilizing the

Nyquist range, we set $\nu_0$ at the edge of the measured optical spectrum. Thus, the trade-off relation may be described with an optical bandwidth of the spectrum $\Delta\nu$ as:

$$c_g \Delta\nu < \frac{f_s}{2}. \tag{S22}$$

The spectral resolution $\delta\nu$ determined by the maximum group delay $\tau_{g,max}$ is described as:

$$\delta\nu = \frac{1}{\tau_{g,max}} = \frac{f_{scan}}{c_g}, \tag{S23}$$

where $f_{scan}$ denotes the scan rate. Finally, the trade-off relation among the scan rate, spectral bandwidth and resolution can be described as:

$$f_{scan} \Delta\nu \frac{1}{\delta\nu} < \frac{f_s}{2}. \tag{S24}$$

**B-2. Phase-controlled FTS with a polygonal scanner**

When using a polygonal scanner as a scanning mirror, the above expressions are modified as follows. The down-converted radio frequency of the spectrum may be described as:

$$f(\nu) = \frac{4l_f N \omega}{\nu_0}(\nu - \nu_0) - \frac{2R\omega^2}{c}\nu t. \tag{S25}$$

All the spectral elements $f(\nu)$ must be in the Nyquist range of the system for avoiding the aliasing effect:

$$0 \leq \frac{4l_f N \omega}{\nu_0}(\nu - \nu_0) - \frac{2R\omega^2}{c}\nu t < \frac{f_s}{2}. \tag{S26}$$

This configuration allows us to use a part of the measurement time in order to avoid the aliasing effect. To consider the duty cycle $D$, we introduce $t_{start}$ and $t_{end}$ ($-\frac{1}{f_{scan}} < t_{start}, t_{end} \leq 0$) for expressing the start and end time of the single interferogram, respectively. The duty cycle may be written as:

$$D = (t_{end} - t_{start})f_{scan}. \tag{S27}$$

Assuming the spectrum spanning from $\nu_0$ to $\nu_0 + \Delta\nu$ mapped on the Fourier plane, $t_{start}$ and $t_{end}$ must satisfy the following inequalities:

$$t_{start} > \frac{c}{2R\omega^2} \frac{1}{(\nu_0 + \Delta\nu)} \left(\frac{4l_f N \omega}{\nu_0} \Delta\nu - \frac{f_s}{2}\right), \tag{S28}$$

$$t_{end} \leq 0. \tag{S29}$$

Note that since the frequency of each spectral element decreases in time, the highest available frequency $\frac{f_s}{2}$ determines the start time $t_{start}$, while the lowest frequency zero the end time $t_{end}$. Here, we consider a case where $t_{end} = 0$ so that we have the longest duty cycle. Inequality S28 can be described as:

$$c_p(t_{start})\Delta\nu + \frac{2R\omega^2}{c}\nu_0 \frac{D}{f_{scan}} < \frac{f_s}{2}. \tag{S30}$$

where $c_p(t_{start}) = \frac{4l_f N \omega}{\nu_0} - \frac{2R\omega^2}{c} t_{start}$. The spectral resolution $\delta\nu$, which is an inverse of the group delay during the

time between $t_{start}$ and $t_{end}$, may be written as:

$$\delta v = \frac{1}{c_p(t_{end})t_{end} - c_p(t_{start})t_{start}} \quad (S31)$$

$$= \frac{f_{scan}}{c_p(t_{start})D}.$$

Finally, the trade-off relation is described as:

$$\frac{f_{scan}}{D}\Delta v \frac{1}{\delta v} + \frac{8\pi^2 R v_0 f_{scan}}{cP^2}D < \frac{f_s}{2}. \quad (S32)$$

Here we express the angular frequency of the scanner as $\omega = \frac{2\pi f_{scan}}{P}$, where P denotes the number of facets of the scanner. This trade-off relation clearly shows that the nonlinear phase delay broadens the down-converted RF spectrum, leading to slight reduction of the efficiency on the trade-off relation. Note that Equation S32 equals to Equation S24 with $D = 1$ and $R = 0$.

While, the inequality S28 gives a constraint on $t_{start}$ caused by the sampling frequency, the geometry of the polygonal scanner also gives another constraint on $t_{start}$ as:

$$t_{start} \leq \frac{-\frac{\pi}{P} + \tan^{-1}\left\{\frac{cl_f N}{R}\left(\frac{1}{v_0} - \frac{1}{v_0 + \Delta v}\right)\right\}}{\omega} \quad (S33)$$

$$\approx -\frac{\pi}{2\pi f_{scan}} + \frac{cl_f NP}{R2\pi f_{scan}}\left(\frac{1}{v_0} - \frac{1}{v_0 + \Delta v}\right).$$

In this case, the duty cycle can be described as:

$$D \leq \frac{1}{2} - \frac{Pcl_f N}{2\pi R}\left(\frac{1}{v_0} - \frac{1}{v_0 + \Delta v}\right). \quad (S34)$$